\documentclass{article}%%%%where rsproca is the template name
\usepackage{float}
\usepackage{graphicx}
\usepackage{amssymb,amsmath}
\usepackage{mathtools}
\usepackage{amsfonts}
\usepackage[utf8]{inputenc}
\inputencoding{utf8}
\usepackage[english]{babel}
\usepackage{upgreek}
\usepackage{setspace}
\usepackage{array}
\usepackage{hyperref}
\usepackage[normalem]{ulem}
\usepackage{color}
\usepackage[normalem]{ulem}

\begin{document}
%\pretitle{\begin{center}\Huge\bfseries % Article title formatting
%\posttitle{\end{center}} % Article title closing formatting
%\author{%
%\textsc{John Smith}\thanks{A thank you or further information} \\[1ex] % Your name
%\normalsize University of California \\ % Your institution
%\normalsize \href{mailto:john@smith.com}{john@smith.com} % Your email address
%\and % Uncomment if 2 authors are required, duplicate these 4 lines if more
%\textsc{Jane Smith}\thanks{Corresponding author} \\[1ex] % Second author's name
%\normalsize University of Utah \\ % Second author's institution
%\normalsize \href{mailto:jane@smith.com}{jane@smith.com} % Second author's email address
%}

\title{Wavefront shaping to improve beam quality: converting a speckle pattern into a Gaussian spot} % Article title
\maketitle

\author{\textbf{A.M. Paniagua-Diaz$^{1,2}$, W.L. Barnes$^{2}$ and J. Bertolotti$^{2}$}
\newline
\begin{center}
\normalsize{$^{1}$Universidad de Murcia, Laboratorio de Optica, Departamento de Fisica, Murcia, Spain\\
$^{2}$University of Exeter, Physics and Astronomy Department , Exeter, EX4 4QL, United Kingdom}
\end{center}
%%%% Subject entries to be placed here %%%%
%\subject{Optics, Physics}
%%%% Keyword entries to be placed here %%%%
%\keywords{wavefront shaping, scattering, speckle, beam quality}

%\textsc{John Smith}\thanks{A thank you or further information} \\[1ex] % Your name
%\normalsize University of Murcia \\ % Your institution
%\normalsize \href{mailto:a.paniagua-diaz@um.es}{a.paniagua-diaz@um.es} % Your email address
%\and % Uncomment if 2 authors are required, duplicate these 4 lines if more
%\textbf{William L. Barnes}\thanks{Corresponding author} \\[1ex] % Second author's name
%\normalsize University of Exeter \\ % Second author's institution
%\normalsize \href{mailto:jane@smith.com}{jane@smith.com} % Second author's email address
}
%\date{\today} % Leave empty to omit a date
%\renewcommand{\maketitlehookd}{%
%\begin{abstract}
%\noindent \blindtext an abstract test% Dummy abstract text - replace \blindtext with your abstract text
%\end{abstract}
%}

%\begin{document}

%%%% Article title to be placed here
%\title{Wavefront shaping to improve beam quality: converting a speckle pattern into a Gaussian spot}

%\maketitle

%\author{%%%% Author details
%A.M. Paniagua-Diaz$^{1,2}$, W.L. Barnes$^{2}$ and J. Bertolotti$^{2}$}

%%%%%%%%% Insert author address here
%\address{$^{1}$Universidad de Murcia, Laboratorio de Optica, Departamento de Fisica, Murcia, Spain\\
%$^{2}$University of Exeter, Physics and Astronomy Department , Exeter, EX4 4QL, United Kingdom}

%%%% Subject entries to be placed here %%%%
%\subject{Optics, Physics}

%%%% Keyword entries to be placed here %%%%
%\keywords{wavefront shaping, scattering, speckle, beam quality}

%%%% Insert corresponding author and its email address}
%\corres{A.M. Paniagua-Diaz\\
%\email{a.paniagua-diaz@um.es}}

%%%% Abstract text to be placed here %%%%%%%%%%%%
\begin{abstract}
A perfectly collimated beam can be spread out by multiple scattering, creating a speckle pattern and increasing the étendue of the system. Standard optical systems conserve étendue, and thus are unable to reverse the process by transforming a speckle pattern into a collimated beam or, equivalently, into a sharp focus. Wavefront shaping is a technique that is able to manipulate the amplitude and/or phase of a light beam, thus controlling its propagation through such  media. Wavefront shaping can thus break the conservation of étendue and, in principle, reduce it. In this work we study how much of the energy contained in a fully developed speckle pattern can be converted into a high quality (low $M^2$) beam, and discuss the advantages and limitations of this approach, with special attention given to the inherent variability in the quality of the output due to the multiple scattering.
\end{abstract}
%%%%%%%%%%%%%%%%%%%%%%%%%%%

%%%%%%%%%% Insert the texts which can accomdate on firstpage in the tag "fmtext" %%%%%

%\begin{fmtext}
\section{Introduction}
Reflective and refractive optics, such as mirrors and lenses, can be used to manipulate light propagation in many ways, but there are limitations in what they can do. For instance any number of mirror and lenses will always conserve the étendue of the system, which quantifies how ``spread out'' a beam is in both area and angle. As a result it is always possible to produce a wide collimated beam from the light coming from a very small source, trading area for angular spread, but one can not create a sharp focus from a wide source that has a large angular spread. There are of course exceptions to étendue conservation: 
%\end{fmtext}

%%%%%%%%%%%%%%% End of first page %%%%%%%%%%%%%%%%%%%%%

\noindent for instance it is trivial to increase it by placing a diffuser in the optical path \cite{aa}. It is also trivial to reduce it by placing a pinhole in the beam, but this comes at the cost of huge energy losses. This works also for perfectly coherent light, where multiple scattering, e.g. by biological tissues, will increase the étendue and results in a seemingly random speckle pattern \cite{1,1a}. 

In recent years wavefront shaping techniques have emerged as a method capable of taking advantage of the properties of multiple elastic scattering, demonstrating the capability of controlling the propagation of light in disordered media \cite{4,5,6} by means of amplitude and/or phase manipulation of the light beam. These techniques were originally proposed to focus light through a scattering material \cite{4}, and have since proved very useful in different fields such as imaging \cite{7,8,9,10,11}, enhancing energy delivery \cite{12,13,14} or cryptography \cite{15,16}. In principle wavefront shaping techniques should be able to completely control the propagation of light in a disordered medium and even reverse the effect of scattering. However in practice this is not possible due to the limited  number of degrees of freedom available in existing wavefront synthesizers or Spatial Light Modulators (SLM). Both the best possible and the expected value for the enhancement achievable with wavefront shaping have been discussed in the literature, together with the effect of several possible limiting factors (e.g. sample stability, phase/amplitude only modulation etc.)  \cite{5,17,18}. 

In this work we experimentally demonstrate that, using wavefront shaping, it is possible to convert a fully developed speckle pattern into a diffraction limited spot, significantly reducing its étendue. We discuss in detail the advantages and limitations of this approach, and in particular how much of the total energy can be placed in the focus, and how this is connected to the field enhancement and its probability distribution.

\section{Converting speckle to a focus}
%\section{Beam Quality Improvement}
Multimode fiber-based lasers are compact and robust, and thus excellent candidates to generate high power. These lasers can achieve high power outputs at very narrow linewidth, which makes them very appealing for different applications \cite{19,20}. However for some spectral regions, in order to achieve high powers, overheating can only be avoided by making use of multimode optical fibers. The problem with multimode fibers is that each mode propagates at a different speed, resulting in a random speckle pattern and hence poor beam quality. Traditional solutions to improve the beam quality involve placing a pinhole or a single mode fiber so as to filter one spot from the speckle pattern, but both are extremely energy inefficient and negate the power gain by employing multimode fibers \cite{27}.
Wavefront shaping can be used to imprint a phase profile on a gaussian beam that compensates for the mode dispersion in the fiber, thus producing a sharp focus at the distal end \cite{21}. As speckle patterns are, despite their appearance, spatially coherent \cite{1}, it should be possible to use wavefront shaping to refocus them. The problem is that such an approach would require both amplitude and phase modulation, with the amplitude modulation able to amplify the signal coming from the areas of low or zero intensity of the speckle pattern, which is outside the capabilities of any realistic wavefront shaping scheme. A complete study of an efficient refocussing of a fully developed speckle pattern, such as the one emerging from a multimode fibre, has not been demonstrated yet, to the best of our knowledge, it is the focus of the present report.

\begin{figure}[h]
\centering
\includegraphics[width=0.8\linewidth]{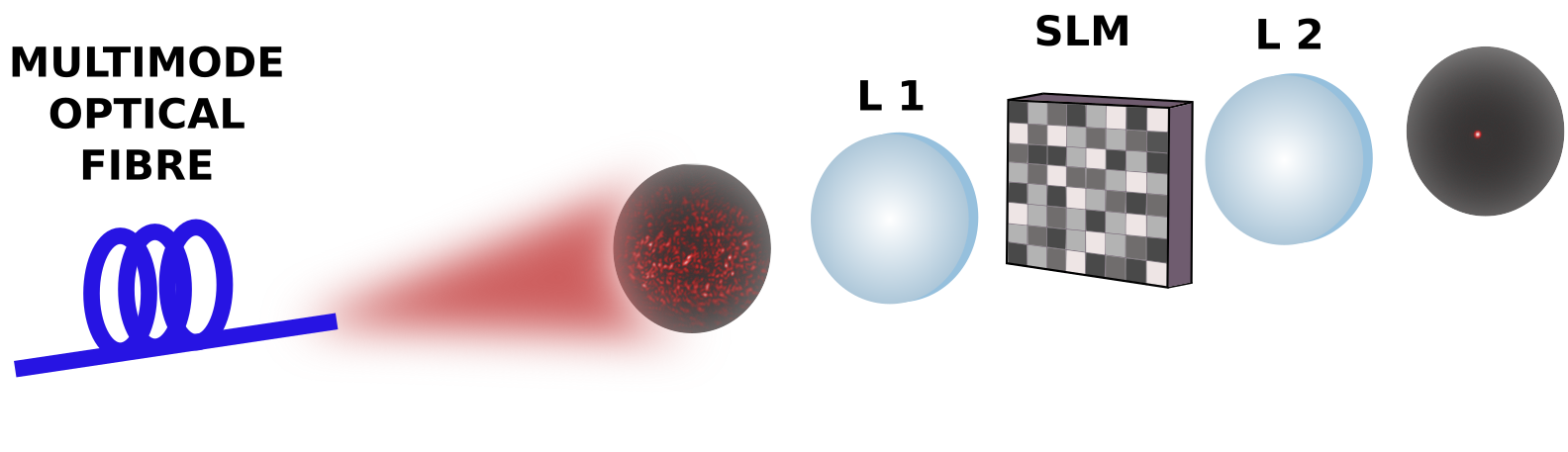}
\caption{Schematic of the experimental setup used to improve the beam quality of a multimode beam by wavefront shaping. The output of the multimode fiber is collimated and sent to the SLM, where the spatial phase distribution of the wavefront is modified in order to compensate for the random scattering of the modes in the fiber and create a collimated beam, which is then focused by a lens.}
\label{fig:Setup}
\end{figure}

\begin{figure}[h]
	\centering
	\includegraphics[width=0.98\linewidth]{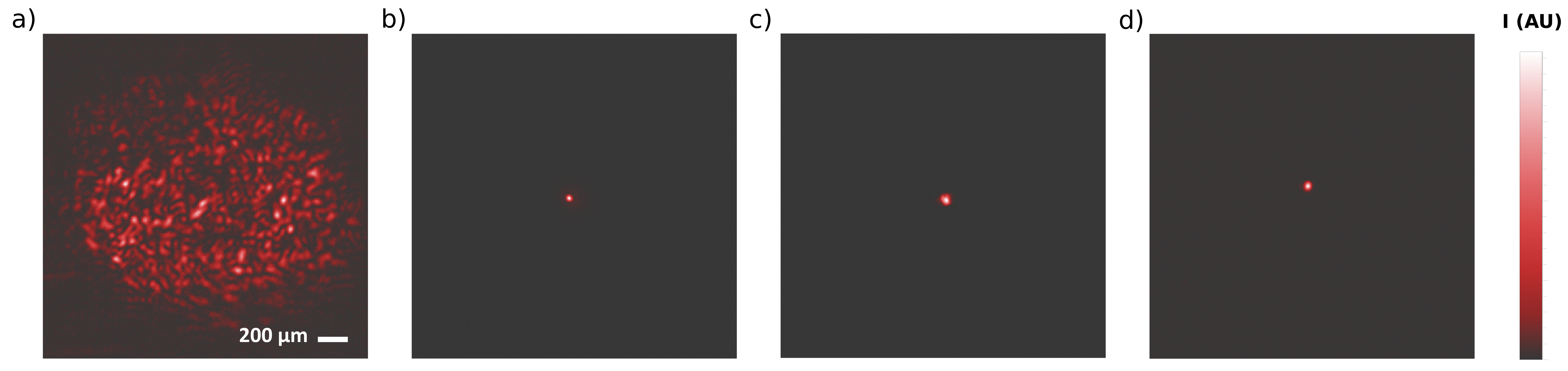}
	\caption{Images of the beam waists at the focal plane of a $750$mm lens under different configurations. a) Beam waist of the output from the multimode optical fiber, showing the speckle pattern of the beam. The beam quality factor is $M^2=59 \pm 5$. %The total power delivered with the HeNe laser used is $1.07$ mW.
b) Beam waist of the output of a single mode fiber. The beam quality factor in this case is $1.4 \pm 0.4$. %The power in this case, using the same laser is $3.3\upmu$W \todo[inline]{TV: Weird sentence, not sure why the second part is needed.}. 
 c)Beam waist of a filtered speckle spot using a $100 \upmu$m diameter pinhole, acting as a simple spatial filter. The beam quality factor in this case is $1.5 \pm 0.3$.% and the power filtered in this mode $4 \upmu$W. 
 d)This picture shows the beam filtered through the pinhole after completion of the wavefront shaping optimization. The beam quality factor in this case is $1.2\pm 0.3$.% and the extrapolated power in the mode, assuming no losses in the SLM, $256 \upmu$W.
 }
\label{fig:Figs}
\end{figure}

A schematic of the experimental setup is shown in figure \ref{fig:Setup}. The experiment was done using a HeNe laser (632.8 nm at 6 mW) coupled to a step-index multimode fiber (core diameter 550 $\upmu$m, NA=0.22) and a Digital Micromirror Device (DMD) as SLM (Vialux DLP Discovery 4100, Texas Instruments). DMDs are amplitude modulators, but they can also be used to modulate phase using holographic techniques such as Lee Holography \cite{28}. We preferred phase modulation over amplitude modulation as phase modulation is more efficient for wavefront shaping \cite{6}. The output of the multimode optical fiber is collimated and sent to the SLM to shape the beam. A second lens collects the modulated beam and focuses it. The total intensity in the focus is used as the metric for the optimization, as widely used in the literature for evaluating the quality of the wavefront shaping technique \cite{5}. We place a $100 \upmu$m diameter pinhole acting as a spatial filter on the focal plane, and compare the amount of light that can be concentrated in that area with and without wavefront shaping (figure~\ref{fig:Figs}).

\noindent %Figure~\ref{fig:Figs}\textit{a} shows the image of the beam waist at the output of the fiber before any optimization is done, where the random interference between the different modes as they propagate through the multimode optical fiber is visible as a speckle pattern. 
To evaluate the beam quality of this and subsequent beams, we use the standardized beam quality factor $M^2$, which compares any beam with an ideal Gaussian diffraction-limited beam \cite{22}:
\begin{equation}
	M^2 = \frac{\pi \omega_0 \theta}{\lambda},
\end{equation}
where $\omega_0$ is the radius at the beam waist (or waist radius), $\theta$ the divergence angle and $\lambda$ the wavelength of the beam. 

\noindent The beam waist at the output of the multimode fiber (figure \ref{fig:Figs}\textit{a}) shows a speckle pattern with a large number of diffraction limited spots. This results in a poor beam quality, with an $M^2$ factor equal to $59 \pm 5$. If we couple the beam to a single mode fiber, we can readily obtain a diffraction-limited spot at the beam waist (figure \ref{fig:Figs}\textit{b}). In this case $M^2 = 1.4 \pm 0.4$, which is very close to the ideal value of $1$ for a diffraction limited Gaussian beam.
However this process is extremely energy-inefficient as, due to the conservation of étendue, only a small fraction of the total energy can be coupled to the single mode fibre. For our setup, less than $0.5$\% of the power could be coupled in this way. The same is true if one uses a pinhole to increase the beam quality: a $100 \upmu$m diameter pinhole produces a good quality beam ($M^2=1.5 \pm 0.3$), at the expenses of energy efficiency, which is again reduced to less than $0.5\%$ (figure \ref{fig:Figs}\textit{c}).

To obtain both a high quality beam and good energy efficiency we use wavefront shaping, i.e. we use a spatial light modulator to imprint a position-dependent phase profile on the wavefront, such that constructive interference is built at one chosen speckle spot, allowing us to deposit most of the available beam energy at that place. In figure \ref{fig:Figs}\textit{d} we show the result of using wavefront shaping techniques to increase the intensity deposited into a diffraction limited spot. In this experiment we achieved enhancement factors of $300 \pm 13$, i.e. the intensity in the optimized spot became 300 times larger than the average intensity of any spot in the multimode beam (Figure \ref{fig:Figs}\textit{a}). 
In this case the improvement of the $M^2$ factor changed from $59 \pm 5$ to $1.2 \pm 0.3$.

\begin{figure}[h]
	\centering
	\includegraphics[width=0.5\linewidth]{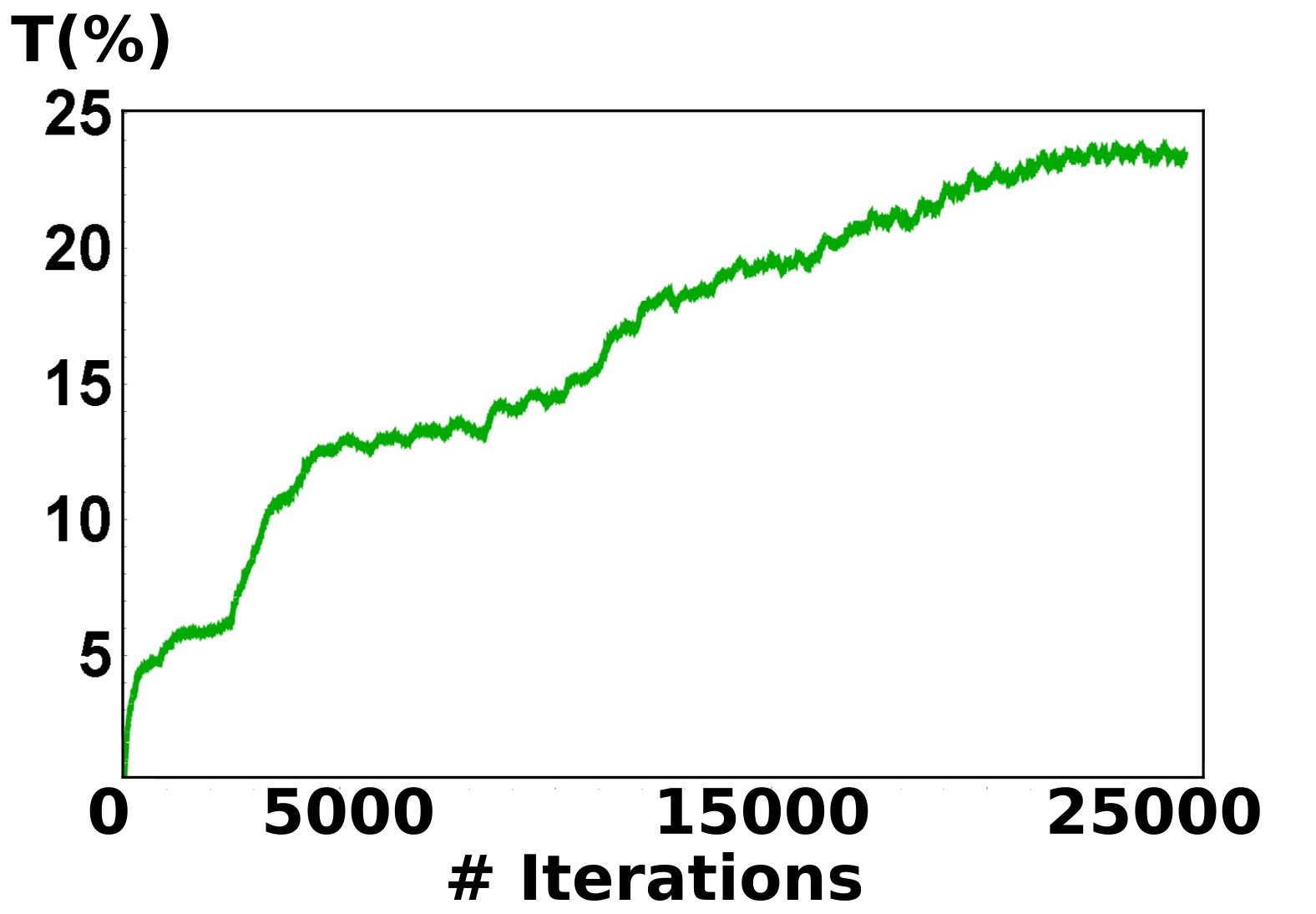}
	\caption{The total intensity transmitted through the pinhole can be seen to increase with the number of iterations of the algorithm until it reaches a plateau at approximately $24\%$ after $24500$ iterations.}
	\label{fig:Increase}
\end{figure}

The plot in figure \ref{fig:Increase} shows the increase in the intensity deposited in the filtered diffraction-limited spot as the algorithm progresses. The power of the filtered mode is normalized by the total power of the unoptimized multimode beam. From this graph we can see that the power deposited in the optimized mode goes from an initial value of the total intensity smaller than $0.5\%$ (when considering a simple pinhole filtering) to a value that varies between $23\%$ and $25\%$ of the total intensity when the transmission through the pinhole is optimized. The maximum speed of this optimization depends on several factors such as the strength of the signal and the type of processing used. Using a DMD as a spatial light modulator, a highly sensitive photodiode and implementing the control algorithm in a Field Programmable Gate Array (FPGA) the whole system can reach optimization rates up to 5kHz, resulting in optimization times of approximately 5 seconds \cite{23}.

\section{Fluctuations of the enhancement factor}
%\section{Theoretical focusing enhancement}
If wavefront shaping was ideal, including the ability to amplify the signal from the low intensity areas, one could reach $100$\% energy efficiency. In practice this is not possible for any realistic scheme, and it is important to quantify both the expected outcome and its fluctuations. To do so we study the optimal enhancement factor that can be achieved under ideal wavefront modulation and illumination, presenting its probability distribution for the first time, as well as the most adverse illumination and the factors that should be taken into account in order to obtain an accurate experimental estimation of the enhancement factor in experimental conditions.

To quantify the optimal focusing efficiency, we can represent the scattering material by a transmission matrix $t$, connecting the input fields of the material to the output fields \cite{24}. The outgoing scattered fields are given by: 
\begin{gather}
\begin{bmatrix}
E_{o1} \\
E_{o2} \\
\vdots \\
E_{oM}
\end{bmatrix} 
=
\begin{bmatrix}
t_{11} & t_{12} &  \dots  & t_{1M} \\
t_{21} & t_{22} &  \dots  & t_{2M} \\
\vdots & \vdots &  \ddots & \vdots \\
t_{M1} & t_{M2} & \dots  & t_{MM}
\end{bmatrix}
\begin{bmatrix}
E_{i1} \\
E_{i2} \\
\vdots \\
E_{iM}
\end{bmatrix} 
\label{trwav}
\end{gather}
\noindent where $E_o$ are the components of the outgoing field, $E_i$ the components of the incident field and $t_{o i}$ are the elements of the transmission matrix connecting both fields. The dimensions of the transmission matrix $M$ are given by the number of propagating modes supported by the sample, and these can be quantified by the number of diffraction limited spots that are in the illuminated area at the entrance or input of the sample \cite{5}. If the task is to focus to one diffraction limited spot (or one output mode), the field contribution to that given spot is: 

\begin{equation}
E_{o f} = \sum_{j}^{N} t_{f j} E_{i j} \, ,
\label{tmatra}
\end{equation}
\noindent where $E_{o j}$ is the field at the chosen output channel. The focusing figure of merit is the enhancement factor, accounting for the intensity at the optimized spot with respect to the average initial intensity \cite{4}. 

\subsubsection*{Optimal focusing enhancement factor}

The focusing enhancement factor, $\eta$, is defined as the ratio between the intensity at the optimized speckle spot and the average intensity at that spot when the wavefront is not optimized \cite{4,18}: 
\begin{equation}
\eta = \frac{I_{f}}{\langle I_{f}\rangle} \, ,
\label{en}
\end{equation}
where $I_{f}$ is the intensity at the optimized speckle spot $f$ and $\langle I_{f}\rangle$ is the ensemble averaged intensity at the same spot over different realizations of disorder when the wavefront is not optimized. 
The following derivation is based on \cite{17,18}. We use eq.~\ref{tmatra} to write the intensity at one speckle spot as:

\begin{equation}
	I_{f}=\mid \sum_{i}^{N} t_{f i} \cdot E_{i} \mid^2 \, .
	\label{intsum}
\end{equation}

\noindent The optimal value the intensity $I_f$ in equation \ref{intsum} can take is determined by the Cauchy-Schwartz inequality: 
\begin{equation}
	\mid \sum_{i}^{N} t_{f i} \cdot E_{i}^{\ast} \mid^2 \leq \sum_{i}^{N} \mid t_{f i} \mid^2 \sum_{i}^{N} \mid E_{i} \mid^2 .
	\label{caus}
\end{equation}
\noindent The maximal intensity we are able to concentrate at one speckle spot is then given by the equality of these two terms. The two terms are equal if the electric field $E_i = C \ t_{f i}^{\ast}$ where $C \in \mathbb{C}$ and in our case is a normalization factor: $C={\left( \sqrt{ \sum_{i}^{N} \mid t_{f i} \mid^2}\right) }^{ -1}$ so that $\sum_{i}^{N}{\mid E_{i}\mid^2 =1}$. We refer to the field satisfying the equality as $\tilde{E_i}$ to indicate it is the optimal field for the desired output. The optimal intensity at the spot $f$ is then given by: 
\begin{equation}
	\tilde{I_{f}}=\sum_{i}^{N} \mid t_{f i} \mid^2 \sum_{i}^{N} \mid \tilde{E_{i}} \mid^2
	\label{iopt}
\end{equation}
\noindent The ideal enhancement factor is then given by (complete derivation in Appendix A):
\begin{equation}
\tilde{\eta} = \frac{\tilde{I}_{f}}{\langle I_{f}\rangle} = \frac{\sum_{i}^{N} \mid t_{f i} \mid^2}{\langle \mid \xi_{f i} \mid^2 \rangle},
\label{idenh}
\end{equation}
where  $t_{f i}$  and $\xi_{f i}$ are elements of the transmission matrix for the optimized and non-optimized fields, respectively. If we average the optimal enhancement factor from equation \ref{idenh}, we retrieve the optimal average enhancement derived originally by Vellekoop \cite{5}:
\begin{equation}
\langle \tilde{\eta}\rangle = N
\label{optimalaverage}
\end{equation}
The expected value of the enhancement factor in eq.~\ref{optimalaverage} is widely used as a reference for the quality of a wavefront shaping experiment, but we do not expect all the experiments to give an enhancement factor equal to this. As both $t_{f i}$  and $\xi_{f i}$ in eq.~\ref{idenh} can be seen as (correlated) random variables, also $\tilde{\eta}$ is a random variable with its own distribution.
Since the real and imaginary parts of the electric field are normally distributed \cite{aa}, the intensity is the sum of the two components squared:$ \mid t_{f i}\mid^2 = \mid \Re[t_{f i}]\mid^2 + \mid \Im[t_{f i}]\mid^2$. Therefore these elements follow a chi-squared distribution $\chi^{2}_2$, or exponential distribution. 
\bigskip

\begin{figure}[ht]
	\centering
	\includegraphics[width=0.8\linewidth]{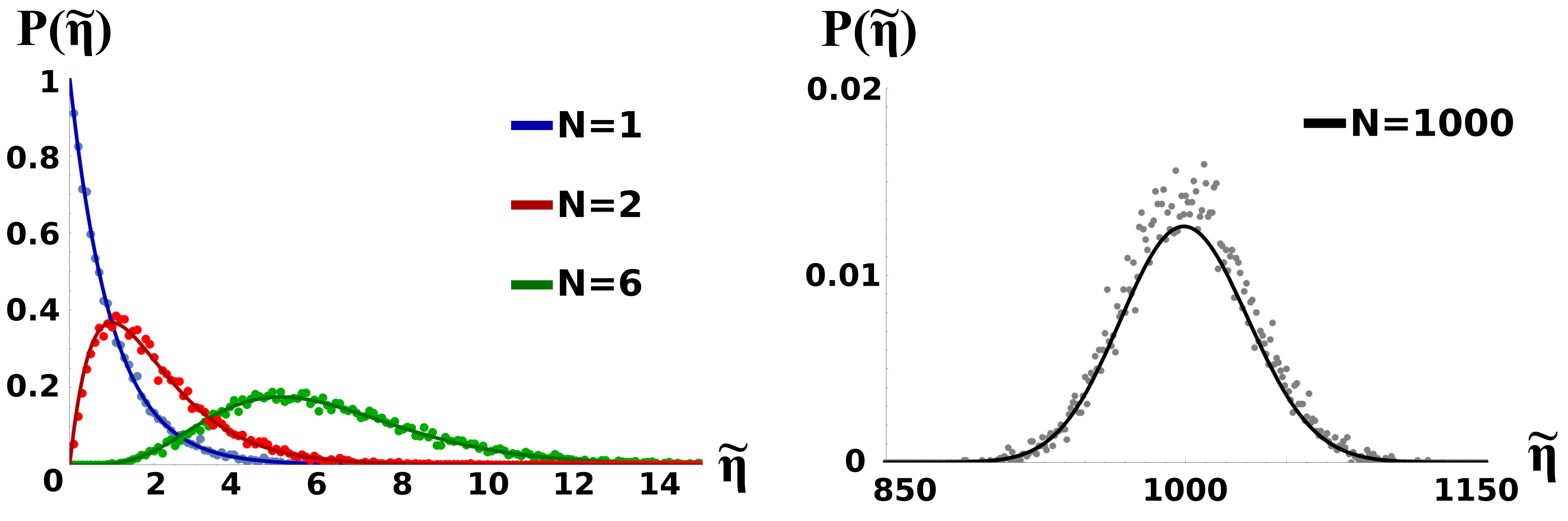}
	\caption{Probability distribution of the enhancement factor for different number of modes $N$. The solid lines represent the probability distribution described in eq.~\ref{prob} and the dots are values obtained by repeatedly performing numerical experiments for the given number of modes. On the left hand side graph we show the widening of the function as the number $N$ increases, converging towards a more symmetric function. The graph on the right hand side shows the function for a large number of independent modes $N=1000$.}
\label{fig:mergedDist}
\end{figure}

The distribution of the enhancement factor is then given by the sum of chi-squared distributed terms over the average value of $\chi^{2}_2$: 
\begin{equation}
P(\tilde{\eta})=\frac{P(\sum_{i}^{N} \mid t_{f i} \mid^2)}{\langle \chi^{2}_2 \rangle} = \frac{P(\sum_{i}^{N} \mid t_{f i} \mid^2)}{2} .
\end{equation}
It follows from the definition of chi-squared distributions that the sum of $N$ terms with $\chi^{2}_2$ distributions is also chi-squared distributed, thus the distribution of the enhancement factor is given by $\chi^{2}_{2 N}$. If we consider the superposition of the two orthogonal polarizations, the resultant distribution is the sum of the two polarizations: 

\begin{equation}
P(\tilde{\eta})=\frac{1}{2} \chi^{2}_{2 N_\bot} + \frac{1}{2}\chi^{2}_{2 N_\parallel} =\frac{1}{2} \chi^{2}_{ (2 N_\bot +2  N_\parallel)} = \frac{1}{2} \chi^{2}_{2 N}
\label{prob}
\end{equation}

\noindent where $N$ is the sum of modes $N=N_{\bot}+N_{\parallel}$ and the factor $\frac{1}{2}$ is given by the denominator in equation~\ref{idenh}, as $\langle \chi^{2}_{2}\rangle =2$. This distribution has an average equal to $N$, as expected from eq.~\ref{optimalaverage}, and a standard deviation $\sqrt{N}$.
In figure~\ref{fig:mergedDist} we show the normalized probability distribution of the ideal enhancement factor together with numerical data for different values of $N$. When $N=1$ we obtain the exponential decay in intensity, whereas when two independent modes are present, the Rayleigh distribution appears as expected \cite{1}. It is of particular interest to note how, for small values of $N$ (easily achievable in the case of small core optical fibers), the enhancement factor can deviate significantly from the mean value. When the number of modes is much larger, as shown in the graph on the right, the distribution is closer to a Gaussian, as a consequence of the central limit theorem.

\subsubsection*{Imperfect enhancement factor}
\noindent In the previous section we discussed the focusing enhancement factor assuming we can synthesize exactly the wavefront that best suits our sample configuration ($\tilde{E}_i$). In practice, perfect control is not possible and therefore the maximal enhancement is reduced by several experimental factors, such as the number of pixels of our SLM, dictating the number of modes we can control, only phase or amplitude modulation or one polarization control, among others. All these experimental imperfections can be captured by the factor $\gamma$, the normalized overlap between the ideal required optimal field and the experimentally synthesized field \cite{5,18}: 
\begin{equation}
\gamma = \sum_{i=1}^{N} \frac{\tilde{E}_{i} E^{*}_{i}}{\sqrt{\tilde{I} I}},
\end{equation}
where $\tilde{I}, I$ are the intensities of the ideal ($\tilde{E}_{i}$) and the experimental (or non-ideal) fields ($E_{i}$), respectively. The experimental synthesized field can then be written as:
\begin{equation}
E_{i}= \gamma \tilde{E}_{i} + \sqrt{1- \mid \gamma \mid^2} \Delta E_{i}.
\label{efield}
\end{equation}
where $ \Delta E_{i}$ is an error term perpendicular to $\tilde{E}_i$ by definition.

\noindent Substituting equation~\ref{efield} into equation \ref{idenh}, we obtain the experimental correction to the ideal enhancement factor. The average experimental enhancement factor is given by:
\begin{equation}
\langle \eta \rangle = \langle \mid \gamma \mid^2 \rangle \langle \tilde{\eta} \rangle + (1+ \langle \mid \gamma \mid^2 \rangle )
\label{enhexp}
\end{equation}
\noindent where $\mid \gamma \mid^2$ is called the fidelity factor. 

Wavefront shaping has generally been used to transform a plane wavefront into a random one, compensating the phase disturbance of the scattering sample it is incident on. However, in our case we are interested in transforming a random wavefront and convert it into a plane wave. Although naively one could think there is no difference between the two options, there are indeed two important differences between using plane wave or speckle illumination. The first thing to take into account is the polarization. When speckle patterns are generated as a consequence of multiple scattering of light, the original polarization is scrambled, resulting in a pattern where both orthogonal polarizations are roughly equally present. If we aim to control a pattern where both polarizations are present, we need to address them independently, given that they interfere independently. If we only control one polarization, the total degrees of freedom one can control are reduced by a factor of two, as per eq.~\ref{prob}. 

The second difference is that a speckle pattern has many areas of zero or low intensity. As realistic wavefront shaping techniques are unable to amplify the intensity, little to no control can be achieved in these areas, and any pixel on the Spatial Light Modulator that happens to sit on a dark spot of the speckle pattern will not contribute much to the final result. This can be accounted for by the fidelity factor \cite{17,18} in the form: 
\begin{equation}
\mid \gamma_{a} \mid^2 = \frac{\overline{A_{i}}^2}{\overline{{A_{i}}^2}} \approx \frac{1}{2}
\end{equation}
where $A_{i}=1+\frac{\delta A_{i}}{\mid \tilde{E}_{i} \mid}$ is the error in amplitude of the field over different incident channels and the over-line represents the spatial average. Details about other contributions to the final enhancement factor are in Appendix B.

Finally, we compare the experimental and expected enhancement factors we obtained. For our system the total number of degrees of freedom is given by approximately the number of speckle spots of the multimode output $N\approx 2000$.  Given that we only control one polarization channel, this is reduced by factor of two, so that the total number of degrees of freedom we can initially control is $N_P=1000$, and therefore the ideal enhancement we could achieve $\tilde{\eta}\approx 1000$. However there are other experimental factors such as amplitude inhomogeneities due to speckle pattern illumination ($\mid \gamma_{a}\mid^2 = \frac{1}{2}$), phase-only modulation ($\mid \gamma_{ph}\mid^2 \approx \frac{\pi}{4}$), temporal decorrelation ($\mid \gamma_{t}\mid^2 \approx 0.9$) and discrete phase modulation ($\mid \gamma_{lee}\mid^2 \approx 0.98$). All this factors accounting for imperfections in the modulation result in an expected enhancement factor $\eta = \tilde{\eta}\mid\gamma\mid^2\approx 340$, in good agreement with the experimental results, indicating that we have established a reliable theoretical framework when speckle patterns are used as illumination for wavefront shaping techniques.

\section{Conclusions}

In this work we demonstrated that wavefront shaping is an effective way to increase the beam quality from a multimode fibre, and that it can increase the $M^2$ factor of a fully developed speckle pattern by just as much as more traditional techniques (from $M^2\sim 60$ to $M^2\sim 1.2$), but with the benefit of significantly smaller energy losses (from a $\sim0.5\%$ to a $\sim25\%$ energy efficiency). We have also analyzed in detail the expected performance of this approach. This combined experimental and theoretical approach demonstrates that wavefront shaping is a viable technique to improve the quality and efficiency of high-power fibre lasers.

\section*{Appendix A. Enhancement factor under ideal illumination}

Following the derivation of the enhancement factor from eq. \ref{en} to eq. \ref{iopt}, the optimal intensity at the spot $f$ is given by:

\begin{equation}
\tilde{I}_{f}=\sum_{i}^{N} \mid t_{f i} \mid^2 \sum_{i}^{N} \mid \tilde{E_{i}} \mid^2 = \sum_{i}^{N} \mid t_{f i} \mid^2,
\end{equation}

\noindent where the last simplification is due to the normalized incident field. 

The non-optimized intensity can be described either as a different field impinging on the same scattering medium ($\mid \sum_{i}^{N} t_{f i} \cdot {E'_{i}} \mid^2 $) or as the same optimized input filed $\tilde{E}_{i}$ impinging onto a different and uncorrelated region of the sample, defined by a different  transmission matrix $\xi_{f i}$. We will use the latter definition for simplicity. In this way, the non-optimized intensity at the spot $f$ is given by: 

\begin{equation}
I_f = \mid \sum_{i}^{N} \xi_{f i} \cdot \tilde{E}_{i} \mid^2 = \sum_{i}^{N} \mid \xi_{f i} \mid^2 \mid \tilde{E}_{i} \mid^2 + \sum_{i}^{N}\sum_{i'\neq i}^{N-1} \xi_{f i} \tilde{E}_{i}^{\ast} \xi_{f i}^{\ast} \tilde{E}_{i} 
\label{nonopt}
\end{equation}

\noindent In order to calculate the enhancement factor, we ensemble average the non-optimized intensity $I_f$ over different realizations of disorder. In this case, the term representing the disorder of the system are the matrix elements $\xi_{f i}$, so we obtain: 

\begin{equation}
\langle I_{f} \rangle= \langle \sum_{i}^{N} \mid \xi_{f i} \mid^2 \mid \tilde{E_{i}} \mid^2\rangle + \langle \sum_{i}^{N}\sum_{i'\neq i}^{N-1} \xi_{f i} \xi_{f i'}^{\ast} \tilde{E}_{i} \tilde{E}_{i'}^{\ast}  \rangle 
\end{equation}

\noindent Assuming different elements of the transmission matrix are uncorrelated, the second part of the sum reduces to zero when averaged over a large number of realizations of disorder, and so the remaining averaged intensity is: 

\begin{equation}
\langle I_{f} \rangle= \sum_{i}^{N} \langle \mid \xi_{f i} \mid^2 \rangle \mid \tilde{E_{i}} \mid^2 = \langle \mid \xi_{f i} \mid^2 \rangle \sum_{i}^{N} \mid \tilde{E_{i}} \mid^2 .
\end{equation}

\noindent given that $\langle \mid \xi_{f i} \mid^2 \rangle$ becomes constant to a very good approximation when it is averaged over a large number of realizations of disorder, the enhancement factor becomes:

\begin{equation}
\tilde{\eta} = \frac{\tilde{I}_{f}}{\langle I_{f}\rangle} = \frac{\sum_{i}^{N} \mid t_{f i} \mid^2 \sum_{i}^{N} \mid \tilde{E}_{i} \mid^2}{\langle \mid \xi_{f i} \mid^2 \rangle \sum_{i}^{N} \mid \tilde{E}_{i} \mid^2} = \frac{\sum_{i}^{N} \mid t_{f i} \mid^2}{\langle \mid \xi_{f i} \mid^2 \rangle}
\label{idealenh}
\end{equation}

\noindent The average value of the enhancement factor is then given by:
\begin{equation}
\langle \eta \rangle= \frac{\langle \tilde{I}_{f}\rangle}{\langle I_{f}\rangle} = \frac{\sum_{i}^{N} \langle \mid t_{f i} \mid^2 \rangle}{\langle \mid \xi_{f i} \mid^2 \rangle} = \frac{N \langle \mid t_{f i} \mid^2 \rangle}{\langle \mid \xi_{f i} \mid^2 \rangle} = N
\label{avenh}
\end{equation}
\noindent where the last simplification is possible given that the matrices $t$ and $\xi$ are uncorrelated and the average over disorder of the absolute value squared is the same for both. Equation \ref{avenh} recovers the results presented in \cite{5}. This result has been broadly used in the literature as the reference optimal value when focusing through scattering media. Although it is an excellent good estimate of what we can hope for, by just looking at the average value we might be missing important information, as discussed previously from the distribution of the enhancement factor, given by equation \ref{idenh} and depicted in figure~\ref{fig:mergedDist}. 

\section*{Appendix B. Fidelity factor}

The fidelity factor is a combination of different independent experimental contributions  \cite{5,18}. The main contributions to decrease the ideal value of the enhancement factor derived in equation \ref{avenh} are phase-only modulation $\mid \gamma_{ph} \mid^2$, temporal decorrelation $\mid \gamma_{t} \mid^2$, among others. In table~\ref{tabular:fidelityfactor} we show the more frequent contributions to the total fidelity factor, without deriving the full expressions. If detailed derivations are of interest, Vellekoop or van Putten thesis are excellent references \cite{17,18}. The total fidelity factor is a linear combination of all the different contributions:

\begin{equation}
\mid \gamma \mid^2 = \mid \gamma_{cont} \mid^2 \mid \gamma_{ph} \mid^2 \mid \gamma_{t}\mid^2 \mid \gamma_{lee} \mid^2  \mid \gamma_{amp} \mid^2. 
\end{equation}

\begin{table}[ht]
	\begin{center}
		\caption{Contributions to the fidelity factor arising from different and independent sources of imperfect modulation}
		{\setstretch{1.5}
			\begin{tabular}[c]{| >{\centering\arraybackslash}m{5cm} | >{\centering\arraybackslash}m{9cm} | } 
				
				\hline
				{\bf SOURCE OF IMPERFECTION} & {\bf CONTRIBUTION} \parbox{0pt}{\rule{0pt}{2ex+\baselineskip}}\\ \hline
				
				{\bf Controlled degrees of freedom \cite{17,18}} &$\mid \gamma_{cont} \mid^2 = \frac{N_C}{N}$\parbox{0pt}{\rule{0pt}{1ex+\baselineskip}} \linebreak $N_C:$ controlled degrees of freedom \parbox{0pt}{\rule{0pt}{2ex+\baselineskip}}\\ \hline
				
				{\bf Phase only modulation \cite{4}} &  $\mid \gamma_{ph} \mid^2\approx \frac{\pi}{4}$   \parbox{0pt}{\rule{0pt}{2ex+\baselineskip}}\\ \hline
				
				{\bf Binary amplitude modulation \cite{6}} & $\mid \gamma_{ba}\mid^2\approx \frac{1}{2 \pi}$ \parbox{0pt}{\rule{0pt}{2ex+\baselineskip}}\\ \hline
				
				{\bf Temporal decorrelation} &  $\mid \gamma_{t}\mid^2 = \tau_C$\parbox{0pt}{\rule{0pt}{1ex+\baselineskip}}  \linebreak  $\tau_C$: decorrelation time \parbox{0pt}{\rule{0pt}{1ex+\baselineskip}}\\ \hline
				
				{\bf Discrete phase modulation \cite{25}} &  $\mid \gamma_{lee} \mid^2 = \left(\frac{sin(\pi/N_{st})}{\pi/N_{st}}\right)^2 $\parbox{0pt}{\rule{0pt}{2ex+\baselineskip}}  \linebreak  $N_{st}:$ number of phase steps  \parbox{0pt}{\rule{0pt}{1ex+\baselineskip}}\\ \hline
				
				{\bf Non-uniform illumination \cite{18}} &$\mid \gamma_{amp} \mid^2 = \frac{\overline{A_{a}}^2}{\overline{{A_{a}}^2}}$\parbox{0pt}{\rule{0pt}{2ex+\baselineskip}} \linebreak  $A_{a}=1+\frac{\delta A_{a}}{\mid E_{a}^{id} \mid}$: amplitude error of the field in channel $a$. The overline is a spatial average\parbox{0pt}{\rule{0pt}{1ex+\baselineskip}}\\ \hline
				
			\end{tabular}
			
			\label{tabular:fidelityfactor}
		}
	\end{center}
\end{table}

\bigskip
\newpage

\section*{Appendix C. Algorithm}
The algorithm used for this experiment is a variation of a partitioning algorithm \cite{26}. In principle one could measure the effect of individual DMD partitions in sequence to obtain the optimal modulation; however, in practice this yields a poor signal-to-noise measurement. Instead we select a random subset of approximately $20\%$ of the partitions and modulate all in phase. Ideally we would like to change each segment individually given that they are independent, so as a trade off, we generate a random phase pattern to start with, changing the phase of approx 20$\%$ of the segments. When the optimization saturates, we reduce the fraction of partitions that are modulated simultaneously, which allows the optimization to keep increasing at approximately constant speed.

\enlargethispage{20pt}
\bigskip

\subsubsection*{Data Access}
The research data supporting this publication are openly available from: \newline

\noindent https://doi.org/10.5281/zenodo.4964616

\subsubsection*{Authors Contributions}
A.M.P.-D. and J.B. developed the idea. A.M.P.-D. performed the experiments and carried out the data analysis. All authors discussed the results and wrote the manuscript.

\subsubsection*{Competing Interests}
The authors declare no competing interests.

\subsubsection*{Funding}
This work was supported by the Leverhulme Trust’s Philip Leverhulme Prize. A. M. Paniagua-Diaz acknowledges support from EPSRC (EP/L015331/1) through the Centre of Doctoral Training in Meta-materials (XM2).

\subsubsection*{Acknowledgments}
We wish to thank Tom Vettenburg for discussions.

%%%%%%%%%% Insert bibliography here %%%%%%%%%%%%%%

\end{document}